\documentclass[conference]{IEEEtran}
\IEEEoverridecommandlockouts
\usepackage{cite}
\usepackage{amsmath,amssymb,amsfonts}
\usepackage{graphicx}
\usepackage{textcomp}
\usepackage{xcolor}

\usepackage{amsmath, amssymb, amsthm,algorithm}
\usepackage[latin1]{inputenc}
\usepackage{xcolor}
\usepackage{latexsym}
\usepackage{graphics,epsfig}
\usepackage{amsfonts}
\usepackage{booktabs}
\usepackage{color}
\usepackage{multirow}
\usepackage[justification=centering]{caption}
\usepackage{graphicx}
\usepackage{adjustbox}
\usepackage{kantlipsum}
\usepackage[caption=false,font=footnotesize]{subfig}
\usepackage{cases}
\usepackage{url}
\usepackage{tabu}
\usepackage{makecell}
\usepackage{soul}

\usepackage{array}
\usepackage{algpseudocode}
\usepackage[justification=centering]{caption}

\usepackage[caption=false,font=footnotesize]{subfig}
\usepackage[T1]{fontenc}
\usepackage{mwe}
\usepackage{subfig}
\usepackage{mathtools,lipsum,cuted}

\usepackage[english]{babel}
\usepackage{verbatim}

\usepackage[english]{babel}

\def\BibTeX{{\rm B\kern-.05em{\sc i\kern-.025em b}\kern-.08em
    T\kern-.1667em\lower.7ex\hbox{E}\kern-.125emX}}
\begin{document}

\title{Increasing Transmission Distance in THz systems with Erasure Coding and Auxiliary Channels}
\author{\IEEEauthorblockN{Cao Vien Phung and Admela Jukan}
\IEEEauthorblockA{Technische Universit\"at Braunschweig, Germany\\
Email: \{c.phung, a.jukan\}@tu-bs.de
}}
\maketitle

\begin{abstract}
We analyze whether the THz transmission distance can be extended with systematic linear network coding (sRLNC) and a low-bitrate additional channel. While various coding techniques have been proposed to mitigate issues of channel quality, and other techniques have used auxiliary channels to enable an effective THz transmission system configuration, their combination is new, and carries potential for significant improvements in transmission quality. Our specific solution is designed to complementing a generic low complexity FEC code by a low complexity erasure code (sRLNC), whereby channel coding is proposed to simultaneously send the native and coded data over two parallel channels, including one main THz channel and one auxiliary channel with comparably lower bitrate. We show theoretically that the proposed system can improve throughput, support higher modulation levels and transfer data over the longer distances with THz communications.
\end{abstract}

\section{Introduction}
The terahertz (THz) frequency band ($0.3$-$10$ THz) is the prime candidate to satisfy the capacity needs of the exponentially growing multi-gigabit mobile data in wireless networks \cite{6248357, Nagatsuma:13,Seeds:15,6050620}. However, sending data with a high modulation level over a long transmission distance in the THz range is a known challenge. The reason is that THz band is quite sensitive to molecular absorption in the atmosphere \cite{7321055, 7444891} resulting in a deteriorated Signal-to-Noise ration (SNR), i.e., a high bit error rate (BER) or symbol error rate (SER) of transmitted data. Significant ongoing research is trying to address this problem, focusing both on system configuration issues and new transciever design engineering.

Two methods towards this end are of interest to us: the usage of auxiliary channels and erasure coding. A Wifi auxiliary channel was used in combination with THz transmission channel for distance estimation of the THz receiver and  air humidity measurements \cite{7444891}. Auxiliary channels have also been deployed in the context of  Ultra-Massive MIMO Systems at THz band, which can address the distance problem and increase the capacity of system \cite{faisal2020ultramassive,8683394}. Erasure codes on the other hand are known to improve the quality of THz signal by protecting them against losses through redundant coded data segments. Currently, there are three main candidate coding schemes for THz communications discussed  in combination with Forward Error Correction (FEC) codes including Turbo, Low Density Parity Check and Polar code \cite{sarieddeen2020overview}. For effective decoding, these codes need to generally be added the number of bits per stream by the larger modulation levels, e.g., $1024$ QAM, which is a challenge in THz systems. It is the combination of coding techniques and auxiliary channels that is new and interesting, as it carries potential for significant improvements in transmission quality. So far, erasure coding used in combination with auxiliary channels was only used in optical fiber networks \cite{7901451} and Free Space Optics (FSO) \cite{7249308}.

In this paper, we propose a combined usage of auxiliary channels and erasure coding in THz transmission, with the aim of addressing the challenges of transmission distance and quality. We utilize a low-complexity sRLNC to complement a generic low complexity FEC, as proposed by the standard \cite{7901451}. The solution of sRLNC and auxiliary channel is a good choice as the source data does not change during sRLNC encoding process in sender, which additionally reduces decoding overhead, to complement a generic low complexity FEC. Furthermore, the auxiliary channel can carry the number of additional bits to support the decoding process at the receiver, so we do not need to increase the modulation level.  The data of each coding generation are simultaneously distributed over two channels, i.e., the native input data of each generation is distributed over the main THz channel, while the redundancy generated during encoding process is distributed over the auxiliary channel with lower or equal bit rate for erasure error correction at the receiver. It should be noted that the auxiliary channel can be configured from either the transmission technologies different with THz main channel or in the same THz technology, but different frequency range to restrict the co-channel interference. We analyze a generic THz system with different configurations, such as transmission distance and modulation formats, to evaluate the transmsission redundancy, the related total code rate and transmission overhead and the transmission rate of additional channel required for a reliable THz system, based on our analytical model. The results show that the two-channel THz system can achieve a comparably higher fault tolerance and throughput with a higher level modulation formats and a longer transmission distance.

\begin{figure*}[!ht]
\centering
\includegraphics[width=2\columnwidth]{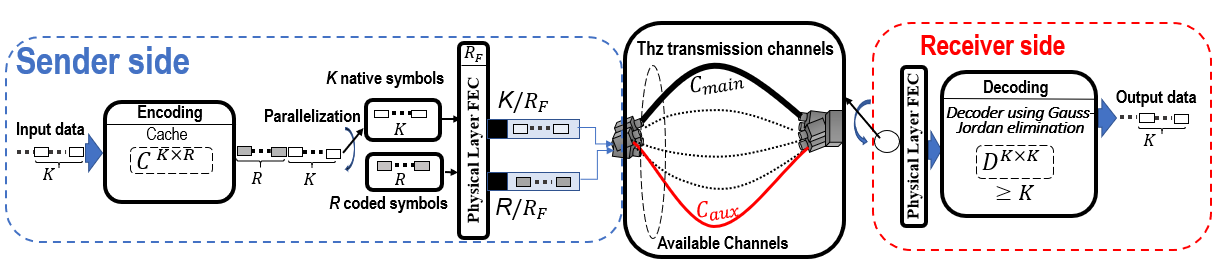}
  \vspace{-0.1cm}
  \caption{Two-channel THz transmission system with sRLNC.}
  \vspace{-0.6cm}
  \label{scenario}
\end{figure*} 

\section{System design}
\begin{table}[t!]
  \centering
  \caption{List of notations.}
  \label{tab:table1}
  \begin{tabular}{ll}
    \toprule
    Notation & Meaning\\
    \midrule
    %$APL_i$ & Acceptable packet loss rate on the $i^{th}$ main channel.\\
    $P_b$ & Residual BER on the main channel.\\
    $P_s$ & Residual SER on the main channel.\\
    $p_e$ & Expected BER on the main channel.\\
    $s$ & Symbol size in bits.\\
    $R_F$ & Code rate of FEC code.\\
    $R_T$ & Total code rate.\\
    $K$ & Generation size in symbols. \\
    $R$ & Redundancy in symbols. \\
    $C_{main}$ & Transmission rate of the main channel.\\
    $C_{aux}$ & Transmission rate of the auxiliary channel.\\
    $\tau_{main}$ & Propagation delay on the main channel.\\
    $\tau_{aux}$ & Propagation delay on the auxiliary channel.\\
    $t_{main}$ & Transmission delay on the main channel.\\
    $t_{aux}$ & Transmission delay on the auxiliary channel.\\
    $T_{main}$ & Total delay on the main channel.\\
    $T_{aux}$ & Total delay on the auxiliary channel.\\
    $d_{main}$ & Transmission distance of the main channel.\\
    $d_{aux}$ & Transmission distance of the auxiliary channel.\\
    %$R_{main}^{T}$ & Transmission rate of the main channel.\\
    %$R_{aux}^{T}$ & Transmission rate of the auxiliary channel.\\
    $\theta$ & Total transmission overhead. \\
     $c_p$ & Propagation speed of light, $c_p=3\cdot 10^{8}$m/s.\\
%	$M$ & Modulation level.\\
    \bottomrule
  \end{tabular}\vspace{-0.5cm}
\end{table}
The reference architecture of two-channel THz system, i.e., including main and auxiliary channels, is shown in Fig. \ref{scenario}. The notations are summarized in Table \ref{tab:table1}.

We propose to utilize sRLNC over both the main and auxiliary channels. Thereby, source data are first encoded to generate coding redundancy, and then distributed over two channels to be simultaneously sent toward receiver. These two channels include one high bitrate main THz channel and one lower or equal bitrate auxiliary channel utilized for transferring original data and coding redundancy, respectively. To restrict the issue of co-channel interference, we need other technologies, e.g., optical fiber, Free Space Optics, Radio Frequency, etc, or the same THz technology but for the different frequency channel, e.g., if the main channel uses THz channel C ($855-890$GHz), then the auxiliary channel utilizes THz channel B ($660-695$GHz) and vice versa, for the auxiliary channel.  The choice of auxiliary channel technology is made depending on the configuration parameters required with regard to the main THz channel, as we later discuss here.

Fig. \ref{scenario} includes three main blocks: Sender, THz transmission channels and receiver. The sender includes two main components: encoder and the FEC coder. The encoder is operated to generate the redundant data from the coding process and the physical FEC code is used to protect data before sending. The THz transmission channels transfer data towards the receiver. Finally, the receiver includes two main components: FEC at physical layer has the function of detecting and correcting error symbols and the decoder has the function of compensating losses which FEC cannot correct. 

As shown in Fig. \ref{scenario}, any input data represented by a long bit stream is split into symbols of $s$ bits each. Any $K$ symbols arrived at the source are first buffered in the coding cache, which is created to temporarily store symbols for the coding process. We define a generation to be a set of symbols coded and decoded. In the figure, each generation has $K$ native symbols. Using sRLNC, the encoding process from the encoder for the redundancy can be performed by multiplying $K$ native symbols with the random coding coefficient matrix $C^{K \times R}$ to generate $R$ redundant symbols for erasure correction at THz receiver. We note that the coding coefficient matrix can be re-used for other generations because they are independently coded and decoded and each coding generation uses a finite field $\mathbb{F}_{2^s}$. The native symbols stay unchanged during the coding process, as are the size of native and coded symbols.

After sRLNC process, $K$ native symbols and $R$ total coded symbols are distributed over $2$ lanes by using round robin, whereby $K$ native symbols and $R$ coded symbols are transferred over one main THz lane and one auxiliary lane, respectively. After the traffic parallelization over $2$ lanes, we apply FEC per lane with the same code rate $R_F$ in the physical layer without consideration of any specific FEC mechanism to protect data before the THz signal transmission. This leads to an increased length of each $K$ native symbols and $R$ coded symbols as $\frac{s \cdot K}{R_F}$ and $\frac{s \cdot R}{R_F}$, respectively. We operate sRLNC in the higher layer (e.g., Ethernet) to complement the function of basic FEC mechanisms that can either fail or yield insufficient performance in the physical layer. Finally, any $K$ native symbols and $R$ coded symbols are sent in parallel over the  main and auxiliary channel, respectively. 

To achieve high bitrate on the main channel, we need to use high-level modulations. Additionally, the auxiliary channel needs to be designed so that its residual BER after FEC is equal to $0$. To this end, the related parameters need to be carefully adjusted, such as modulation level, transmission distance, etc., on both channels to avoid large buffering at the receiver. To keep the transmission rate of auxiliary channel as lowest as possible, which significantly simplifies the design with free residual BER after FEC mechanism at the receiver, the transmission distance of auxiliary channel should be smaller than the transmission distance of main channel (this is proved later in Eq. \eqref{rateaux}).

At the receiver, by using round robin, we collect native and coded symbols from both channels. FEC mechanism at the receiver has the function of detecting error symbols and the receiver will remove them when FEC mechanism cannot correct them. However, the lost symbols from the same generation will be compensated by the sRLNC redundancy on the auxiliary channel. The receiving buffer is necessary but is always a big challenge in high speed systems  to temporally  store all arriving data and we define a decoding window to serially decode each individual generation at the decoder $D^{K \times K}$, whereby each generation needs at least $K$ out of $K+R$ symbols for the decoding process. The sRLNC decoding process is performed by Gauss-Jordan elimination at the decoder and we can only recover original symbols of a generation when we collect more than or equal to the number of symbols of that generation. %\hl{After any generations are completely decoded, they are moved to the higher layers.}

\section{Analysis}
In this section, we provide the analysis of the system presented,  by considering both sRLNC  and FEC code for the total transmission overhead and total code rate. Our main assumptions of the paper are as follows: 
\begin{itemize}
\item The residual BER of auxiliary channel after FEC and co-channel interference between main and auxiliary channels are always equal to $0$.
\item  The bitrate of additional channel is lower than or equal to that of main channel.
\item  The difference of total delay for the same generation between related main and auxiliary channels is equal to $0$.
\item  The redundant data is always sufficient for succesfull decoding.
\end{itemize} 

\subsection{Code rate and transmission overhead}\label{redundancy}
Without loss of generality, the main and auxiliary channels have the minimal Hamming distance of a generic FEC code with a code rate $R_F$ to be  $\Delta^K_{min}=\frac{K \cdot s}{R_F}-(K \cdot s)$ and $\Delta^R_{min}=\frac{R \cdot s}{R_F}-(R \cdot s)$, respectively. Since we assume the residual BER and co-channel interference are equal to $0$, the Hamming distance is only considered for the main channel:

\begin{equation}\label{ham}
\Delta_{min} =\frac{K \cdot s}{R_F}-(K \cdot s) \equiv  \Delta^K_{min}
 \end{equation}. Therefore, the number of erroneous bits that can be corrected by FEC for the main channel is calculated as:
 
 %\begin{equation}\label{t_e}
 %\forall x \in N_M:\text{      }  t_e=\Delta_{min}-1
 %\end{equation}
\begin{equation}\label{t_k}
 \left\{\begin{matrix}
t_k=\frac{\Delta_{min}-2}{2} &, \; \Delta_{min} \; is \; even \\ 
t_k=\frac{\Delta_{min}-1}{2} &, \; \Delta_{min} \; is \; odd
\end{matrix}\right.
 \end{equation}
 
The expected residual BER $P_b$ in Eq. \eqref{eqBER} and residual SER $P_s$  in Eq. \eqref{eqser} after FEC performed by the receiver with expected bit error probability $p_e$ of main THz channel are
  \begin{equation}\label{eqBER}
  P_b = \frac{(K \cdot s \cdot p_e) - (R_F \cdot t_k)}{K \cdot s}
 \end{equation} 
 \begin{equation}\label{eqser}
  P_s=1-(1-P_b)^{s}
 \end{equation}. Note that $P_b \leq 0$ means all bit errors of the main channel are totally
corrected by FEC, i.e., if $P_b < 0$, then assigning $P_b =0$. Thus, with residual SER $P_s$ of the main channel in Eq. \eqref{eqser}, the decoding process needs at least $R=P_s \cdot K$ additional symbols sent over the auxiliary channel, i.e.,
 \begin{equation}\label{RHch}
R\geq P_s\cdot K
  \end{equation}. As a result, with sRLNC redundancy $R$ and FEC code rate $R_F$, the total code rate $R_T$ of both FEC and sRLNC can be calculated as  \begin{equation}\label{RT}
  R_T=\frac{K}{\frac{K+R}{R_F}}=\frac{R_F \cdot K}{K+R}
 \end{equation}. Based on Eq. \eqref{RT}, the total transmission overhead $\theta$ of $2$-channel THz system can be given as
 \begin{equation}\label{overhead}
 \theta = 1-R_T
 \end{equation}

\subsection{Transmission rate}\label{sysana}
Eq. \eqref{Tmain} and Eq. \eqref{Taux} are the total arriving time interval of $K$ native symbols sent on the main channel and $R$ redundant symbols sent on the auxiliary channel, respectively, at the receiver \begin{equation}\label{Tmain}
T_{main}=t_{main}+\tau_{main}=\frac{K \cdot s}{R_F \cdot C_{main}} + \frac{d_{main}}{c_p}
 \end{equation}
 \begin{equation}\label{Taux}
T_{aux}=t_{aux}+\tau_{aux}=\frac{R \cdot s}{R_F \cdot C_{aux}} + \frac{d_{aux}}{c_p}
 \end{equation}
 , where $t_{main}$ and $t_{aux}$ are the transmission delay of $K$ native symbols and $R$  additional symbols, respectively, $\tau_{main}$ and $\tau_{aux}$ stand for the propagation delay of the main channel and  auxiliary channel, respectively, $C_{main}$ and $C_{aux}$ denote a transmission rate of the main channel and auxiliary channel, respectively, $d_{main}$ and $d_{aux}$ are the transmission distance between sending and receiving THz antenna of the main channel and auxiliary channel, respectively, and $c_p=3\cdot 10^8m/s$ is the propagation speed of light.  
 
 With Eq. \eqref{Tmain} and Eq. \eqref{Taux}, in order to avoid a very large buffer, the THz system should be configured so that the difference of arriving time of symbols for the same generation between the main and auxiliary channels is equal to $0$, i.e., $T_{main}=T_{aux}$. Therefore, the transmission rate of an auxiliary channel considered as a function of redundancy and transmission distance between main and auxiliary channels is calculated as \begin{equation}\label{rateaux}
 C_{aux}=\frac{R \cdot s\cdot c_p\cdot C_{main}}{R_F \cdot C_{main} \cdot (d_{main}-d_{aux})+c_p \cdot K \cdot s}
 \end{equation}
Note that $C_{aux} \geq 0$. If $C_{aux}=0$, then we do not need to establish an auxiliary channel to send coding redundancy. With Eq. \eqref{rateaux}, we see that to achive the lowest transmission rate of auxiliary channel, $d_{aux}\ll d_{main}$. Based on Eq. \eqref{rateaux}, we can derive a constraint $R_F \cdot C_{main}(d_{main}-d_{aux})+c_p \cdot K \cdot s>0$, i.e., the transmission distance of auxiliary channel should be as \begin{equation}\label{dlimit}
d_{aux} < \frac{K \cdot s \cdot c_p}{R_F \cdot C_{main}} + d_{main}
 \end{equation}
 
 %%%%%%%%%% Results section
 
 \section{Numerical results}\label{Nume}
 For numerical results, THz channel B ($660-695GHz$) and THz channel C ($855-890GHz$)\cite{7444891}  are simulated by assuming the raised cosine filter with roll-off factor $0.4$ and used omni-directional antennas and the expected BER $p_e$ occurring on any main channel is collected in \cite{7444891}. Each symbol has $8$ bits.
 
\begin{figure}[t]
\centering
\includegraphics[width=0.95\columnwidth]{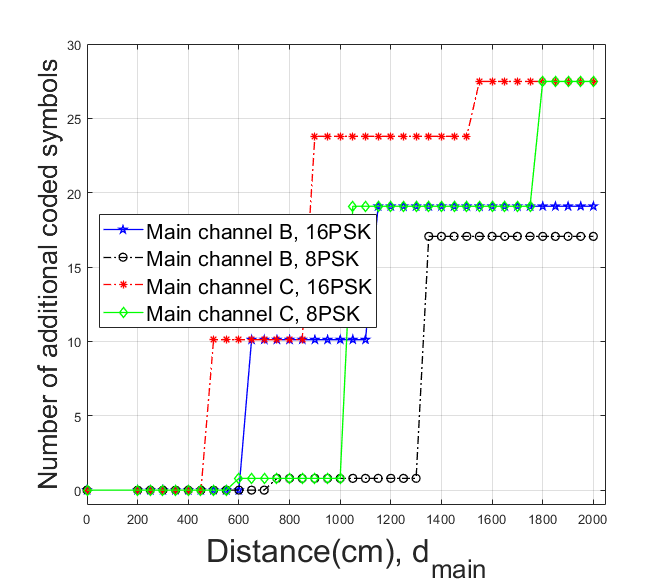}
\caption{Coding overhead vs. transmission distance of main channel B or channel C, $d_{main}$.}
\label{Redundancy}
\vspace{-0.3cm}
\end{figure}

 Fig. \ref{Redundancy} shows the coding overhead (transmission redundancy) distributed over the auxiliary channel for four system configurations. The first and second one use a main channel B with 16PSK and 8PSK, respectively, while the third and fourth one use a main channel C with 16PSK and 8PSK, respectively.  The redundancy $R$ is given by Eq. \eqref{RHch}. The FEC code rate $R_F=0.8$ is used for all channels. The total number of input symbols is set to $K=30$ symbols (\emph{generation size}). The Hamming distance given by Eq. \eqref{ham} is $\Delta_{min}=60$, then with Eq. \eqref{t_k} the number of erroneous bits that can be corrected by FEC  is $t_k=\frac{60-2}{2}=29$ bits. The transmission distance of main channels is configured in the interval $[200,2000cm]$. In this paper, we configure the auxiliary channel so that its residual BER and co-channel interference are approximately equal to $0$. The bitrate of main channel should be configured so that it is higher than or equal to the auxiliary channel.
 
 We can observe in Fig. \ref{Redundancy} that the larger the transmission distance or modulation level, the higher the error bit rate; the redundancy increases with an increasing transmission distance and an increasing modulation level. In addition, the main channel C needs more additional data than the main channel B because channel C incurres more erroneous bits. For main channel B using 16PSK and 8PSK, the sender does not send any coding data when $d_{main}<650cm$ and $d_{main} <750cm$, respectively. For main channel C using 16PSK and 8PSK, the sender does not send any coding data when $d_{main}<500cm$ and $d_{main} <600cm$, respectively. In such cases, the auxiliary channel is unnecessary. However, the additional channel can be required to provide the coded data, when $d_{main} \geq 650cm$ for the main channel B with 16PSK, e.g., $R=11$ coded symbols at $d_{main}=800cm$, $d_{main} \geq 750cm$ for the main channel B with 8PSK, e.g., $R=1$ coded symbols at $d_{main}=1200cm$, $d_{main} \geq 500cm$ for the main channel C with 16PSK, e.g., $R=28$ coded symbols at $d_{main}=1800cm$, and $d_{main} \geq 600cm$ for the main channel C with 8PSK, e.g., $R=20$ coded symbols at $d_{main}=1400cm$.

\begin{figure}[t]
\centering
\includegraphics[width=0.95\columnwidth]{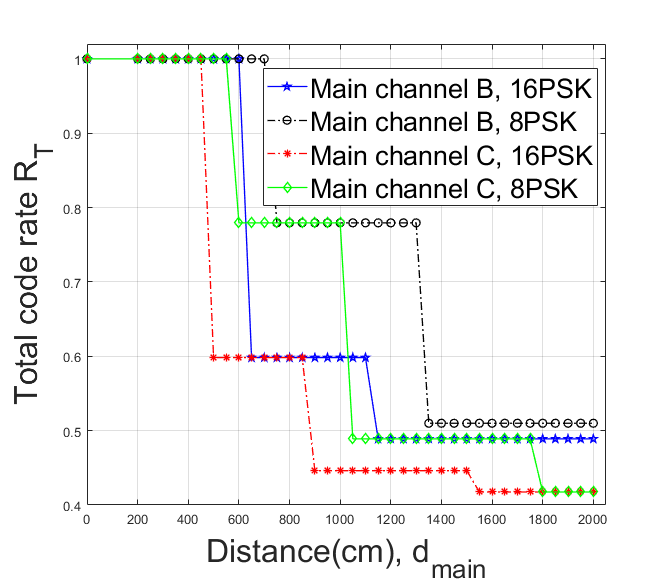}
\caption{Total code rate $R_T$ vs. transmission distance of main channel B or channel C, $d_{main}$.}
\label{Totalcoderate}
\vspace{-0.3cm}
\end{figure}

 Next, based on Eq. \eqref{RT} we show the total code rate for THz transmission with sRLNC and FEC code rate in Fig. \ref{Totalcoderate}. Based on the analysis in Fig. \ref{Redundancy}, we see that the total code rate decreases with an increasing transmission distance and modulation level, and the total code rate of system with main channels C is lower than that of system with main channels B. The total code rate can be in the interval $0.489 \leq R_T \leq 1$ for the system of main channel B with 16PSK, $0.510 \leq R_T \leq 1$ for the system of main channel B with 8PSK, $0.418 \leq R_T \leq 1$ for the system of main channel C with both 16PSK and 8PSK.

\begin{figure}[t]
\centering
\includegraphics[width=0.95\columnwidth]{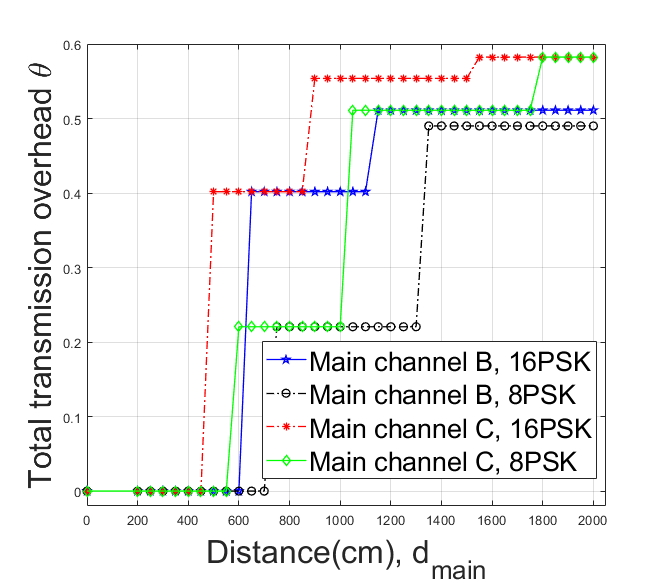}
\caption{Total transmission overhead $\theta$ vs. transmission distance of main channel B or channel C, $d_{main}$.}
\label{Totaltransoverhead}
\vspace{-0.3cm}
\end{figure}

Fig. \ref{Totaltransoverhead} presents the total transmission overhead based on Eq. \eqref{overhead}. Based on the analysis in Fig. \ref{Redundancy}, we see that the total transmission overhead increases with an increasing transmission distance and modulation level, and the total transmission overhead of system with main channels C is higher than that of system with main channels B.  The total transmission overhead can be in the interval $0 \leq \theta \leq 0.511$ for the system of main channel B with 16PSK, $0 \leq \theta \leq 0.490$ for the system of main channel B with 8PSK, $0 \leq \theta \leq 0.583$ for the system of main channel C with both 16PSK and 8PSK.

\begin{figure}[t]
\centering
\includegraphics[width=0.95\columnwidth]{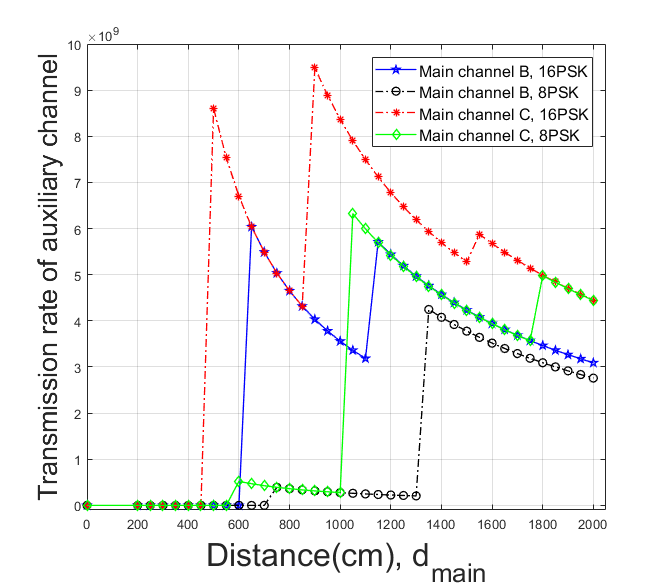}
\caption{Transmission rate of auxiliary channel vs. transmission distance of main channel B or channel C, $d_{main}$, when $d_{aux}=1.5$m.}
\label{rate_aux}
\vspace{-0.3cm}
\end{figure}

Fig. \ref{rate_aux} shows the transmission rate of auxiliary channel required for a reliable THz system to avoid a large receiving buffer size. Here,  we fix the transmission distance of auxiliary channels to $d_{aux}=1.5m$ (note that we should set $d_{aux}$ based on Eq. \eqref{dlimit} so that the transmission rate of auxiliary channel $C_{aux}$ is not negative), $c_p=3 \cdot 10^8 m/s$ as the propagation speed of light, and provide the transmission rate of main channel to be $C_{main}=25GBd/s=2\cdot 10^{11} \cdot L$ \cite{7444891}, where $L$ is the number of bits per symbol for a certain modulation level. The remaining configuration parameters are the same as configured for Fig. \ref{Redundancy}. The transmission rate of auxiliary channels is calculated with Eq. \eqref{rateaux}. $C_{aux}=0$ means there is no need for the auxiliary channel because FEC can correct all error bits from the main channel. Since the behaviors of four kinds of THz system configuration is quite similar, we choose to explain an example of 2-channel THz system with a transmission rate $C_{aux}$ of auxiliary channel supporting for the main channel B with 16PSK. At first $C_{aux}$ increases from $0$ bps to $6.037\cdot 10^9$ bps with $600cm \leq d_{main} \leq 650cm$, then gradually decreases from $6.037\cdot 10^9$ bps to $3.186\cdot 10^9$ bps with $650cm \leq d_{main} \leq 1100cm$, increases from $3.186\cdot 10^9$ bps to $5.711\cdot 10^9$ bps with $1100cm \leq d_{main} \leq 1150cm$ and finally decreases from $5.711\cdot 10^9$ bps to $3.091 \cdot 10^9$ bps with $1150cm \leq d_{main} \leq 2000cm$. The reason of increasing $C_{aux}$ can be explained that those are the increasing amount of additional redundant symbols required, e.g., from $0$ coded symbol at $d_{main}=600cm$ to $11$ coded symbols at $d_{main}=650cm$ and from $11$ coded  symbol at $d_{main}=1100cm$ to $20$ coded symbols at $d_{main}=1150cm$ in Fig. \ref{Redundancy}. The gradual decreasing of $C_{aux}$ is because the considering amount of redundancy sent is constant, e.g., $11$ and $20$ coded symbols for $d_{main}$ intervals $[650cm,1100cm]$ and $[1150cm,2000cm]$, respectively, and at the same time the propagation delay $\tau_{main}$ of main channel 16PSK increases with $d_{main}$ leading to a decreasing transmission rate $C_{aux}$, which has a fixed distance $d_{aux}=150cm<<d_{main}$.  Finally, we conclude that since BER increases with an increasing modulation level of the main THz channel, i.e., increasing the amount of redundancy, the transmission rate of auxiliary channel increases when increasing the modulation level. Additionally, the results confirm our statement in Eq. \eqref{dlimit} that when $d_{aux}<<d_{main}$ the transmission rate of auxiliary channel can be the lowest, whereby in order to reduce the buffering overhead, the native and coded data of the same generation will arrive at the receiver at the same time.

Considering Fig. \ref{rate_aux} again, to restrict the issue of co-channel interference, the auxiliary channel needs to be used the technologies different from the THz main channel. For the main channel B with 16PSK, the auxiliary channel needs $3.091 Gbps\leq C_{aux} \leq 6.037 Gbps$ at $200cm \leq d_{main} \leq 2000cm$ referred to FSO technology with interval $[2.5;10]Gbps$ \cite{tutao}. For the main channel B using 8PSK, the auxiliary channel needs $205.3Mbps \leq C_{aux} \leq 391.7Mbps$ at $750cm \leq d_{main}\leq 1300cm$ referred to $802.11$n WLANs technology with maximum data rate of $600Mbps$  \cite{8422697}, and the auxiliary channel needs $2.759 Gbps \leq C_{aux} \leq 4.246Gbps$ at $1350cm \leq d_{main} \leq 2000cm$ referred to FSO technology. For the main channel C with 16PSK, the auxiliary channel needs $4.319 Gbps\leq C_{aux} \leq 9.482 Gbps$ at $200cm \leq d_{main} \leq 2000cm$ referred to FSO technology. For the main channel C with 8PSK, the auxiliary channel needs $277Mbps \leq C_{aux} \leq 520.6Mbps$ at $ 600cm \leq d_{main}\leq 1000cm$ corresponding to $802.11$n WLANs technology of $600Mbps$  \cite{8422697}.  The auxiliary channel needs $3.567 Gbps \leq C_{aux} \leq 6.323 Gbps$ at $1050cm \leq d_{main} \leq 2000cm $ corresponding to FSO technology.

\begin{figure}[t]
\centering
\includegraphics[width=0.95\columnwidth]{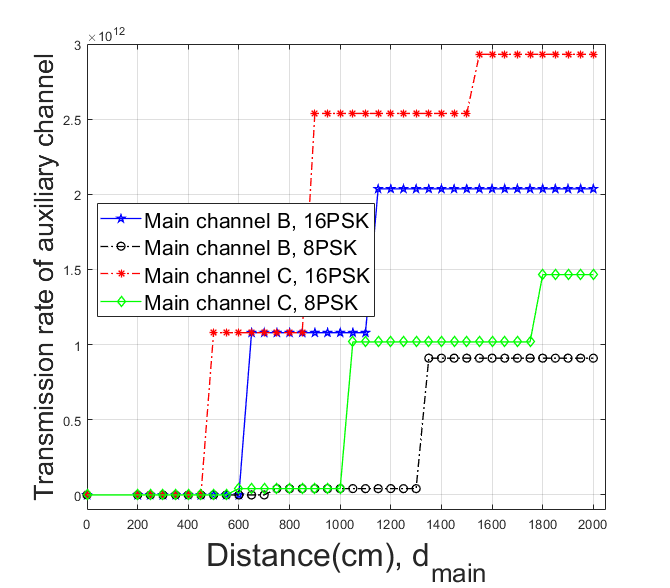}
\caption{Transmission rate of auxiliary channel vs. transmission distance of main channel B or channel C, $d_{main}$, when $d_{main}=d_{aux}$.}
\label{rate_aux_dmainequalaux}
\vspace{-0.3cm}
\end{figure}

Next, as similarly analyzed in Fig. \ref{rate_aux} we show the transmission rate of auxiliary channel for the case of $d_{main}=d_{aux}$ in Fig. \ref{rate_aux_dmainequalaux}. Since $d_{main}=d_{aux}$, based on Eq. \eqref{rateaux} the transmission rate of auxiliary channel in this case is a function of coding redundancy $R$ sent over the additional channel but independent with a transmission distance of the main and auxiliary channel, $d_{main}$ and $d_{aux}$. On the other hand, we see that since the transmission distance of auxiliary channel in Fig. \ref{rate_aux_dmainequalaux} is set to be larger than that in Fig. \ref{rate_aux} ($d_{aux}=1.5m$), the transmission rate of auxiliary channel in Fig. \ref{rate_aux_dmainequalaux} is higher, confirmed our statement from Eq. \eqref{rateaux}. For the main channel B with 16PSK, the auxiliary channel needs $1.079 Tbps\leq C_{aux} \leq 2.036 Tbps$ at $200cm \leq d_{main} \leq 2000cm$ referred to THz technology. For the main channel C with 16PSK, the auxiliary channel needs $1.079 Tbps\leq C_{aux} \leq 2.932 Tbps$ at $200cm \leq d_{main} \leq 2000cm$ referred to THz technology. The auxiliary channel using THz technology should comply with the rules as follows: If the main channel B with frequency range ($660-695GHz$) is used, then channel C with frequency range ($855-890GHz$) is used for the auxiliary channel to restrict the co-channel interference, and vice versa. For the main channel B with 8PSK, the auxiliary channel needs $C_{aux}=42.17Gbps$ at $750cm \leq d_{main} \leq 1300cm$ corresponding to optical fiber technology today, and $C_{aux}=0.910Tbps$ at $1350cm \leq d_{main} \leq 2000cm$ referred to THz technology. For the main channel C with 8PSK, the auxiliary channel needs $C_{aux}=42.2 Gbps$ at $600cm \leq d_{main} \leq 1000 cm$ referred to optical fiber technology and $1.018 Tbps \leq C_{aux}\leq 1.466 Tbps$ at $1050 cm \leq d_{main} \leq 2000cm$ referred to THz technology.

With the analysis in Fig. \ref{rate_aux} and Fig. \ref{rate_aux_dmainequalaux}, we suggest $d_{aux}\ll  d_{main}$ to avoid setting a very high transmission rate of auxiliary channel, which is harder to have a free residual BER after FEC on it. In addition, the technology with high-speed rate is more complicated in the establishment.
\section{Conclusion}
In this paper, we propose a hybrid THz transmission system with low complexity sRLNC complemented for a generic low complexity FEC code, whereby the native data are sent over the main THz channel, while the redundancy is transferred over the additional channel with different frequency ranges of wireless technologies or using fiber technologies. Our analysis evaluates and compares the THz system with different configurations in term of total code rate of FEC and sRLNC, total transmission overhead, coding redundancy, the transmission rate of auxiliary channels required for a certain reliability transmission. The results show that the main channel can send data over longer distances with a high-level modulation, when the additional channel with lower or equal transmission rate supports sending coded symbols for the erasure error correction at the receiver. Our analysis showed that in order to keep the transmission rate of auxiliary channel relatively small, we should set its transmission distance to be much smaller than that of main channel, e.g., $d_{aux}=1.5m$. As a consequence, its transmission rate would fall into Mbps area, which can be cost-effectively realized with FSO or even WiFi technology.  To avoid a very large receiving buffer, all symbols of the same generation will arrive at the receiver at the same time, which we showed to be possible with fixed transmission distance of the auxiliary channel, whereby depending on parameters configured on the main channel, the transmission rate of the auxiliary channel can be calculated such that all symbols of the same generation arrive at the receiver at the same time.
\section*{Acknowledgment}
This work was partially supported by the DFG Project Nr. JU2757/12-1, "Meteracom: Metrology for parallel THz communication channels."
\bibliographystyle{IEEEtran}
\bibliography{nc-rest}

\end{document}